\documentclass[12pt,preprint]{aastex}

\def \kms {$\,$km s$^{-1}$}

\def \mJyb {$\,$mJy$\,$beam$^{-1}$}
\def \h{$^{\rm h}$}
\def \m{$^{\rm m}$}
\def\sec{\hbox{$^{\rm s}$}}

\def \kms {$\,$km s$^{-1}$}

\def \mJyb {$\,$mJy$\,$b$^{-1}$}
\def \arcsec {$^{\prime\prime}$}
\def \arcmin {$^{\prime}$}
\def \arcdeg {$^{\circ}$}
\def \kms {km s$^{-1}$ }

\def \msol {$\,$M$_{\odot}$ }
\def \kms {$\,$km s$^{-1}$}

\def \mJyb {$\,$mJy$\,$b$^{-1}$}

\long\def\symbolfootnote[#1]#2{\begingroup%
\def\thefootnote{\fnsymbol{footnote}}\footnote[#1]{#2}\endgroup}

\shorttitle{Quasi-periodic H$_2$CO Flares in IRAS$\,$18566+0408}
\shortauthors{Araya, et al.}

\begin{document}


\title{Quasi-Periodic Formaldehyde Maser
Flares in the Massive Protostellar Object IRAS$\,$18566+0408}

\renewcommand{\thefootnote}{\fnsymbol{footnote}}

\author{E. D. Araya\altaffilmark{1,2,3},
P. Hofner\altaffilmark{4,5,2},
W. M. Goss\altaffilmark{2},
S. Kurtz\altaffilmark{6},
A. M. S. Richards\altaffilmark{7},\newline 
H. Linz\altaffilmark{8},
L. Olmi\altaffilmark{9,10}, \&
M. Sewi{\l}o\altaffilmark{11}}

\altaffiltext{1}{Physics Department, Western Illinois University, 
1 University Circle, Macomb, IL 61455, USA}
\altaffiltext{2}{National Radio Astronomy Observatory, P.O. Box 0, Socorro, NM 87801, USA}
\altaffiltext{3}{Department of Physics and Astronomy, MSC07 4220,
University of New Mexico, Albuquerque, NM 87131, USA}
\altaffiltext{4}{Max--Planck--Institut f\"ur Radioastronomie, Auf dem H\"ugel 69, 53121 Bonn, Germany}
\altaffiltext{5}{New Mexico Institute of Mining and Technology,
Physics Department, 801 Leroy Place, Socorro, NM 87801, USA}
\altaffiltext{6}{Centro de Radioastronom\'{\i}a y Astrof\'{\i}sica, 
Universidad Nacional Aut\'onoma de M\'exico,
Apdo. Postal 3-72, 58089, Morelia, Michoac\'an, Mexico}
\altaffiltext{7}{Jodrell Bank Centre for Astrophysics, 
School of Physics and Astronomy,
University of Manchester, Manchester M13 9PL, UK } 
\altaffiltext{8}{Max--Planck--Institut f\"ur Astronomie, K\"onigstuhl 17, D--69117 Heidelberg, Germany}
\altaffiltext{9}{University of Puerto Rico at Rio Piedras, Physics
Department, P.O. Box 23343, San Juan, PR 00931, Puerto Rico, USA.}
\altaffiltext{10}{INAF, Osservatorio Astrofisico di Arcetri, 
Largo E. Fermi 5, I-50125, Firenze, Italy}
\altaffiltext{11}{Space Telescope Science Institute, 3700 San Martin Drive, 
Baltimore, MD 21218, USA}

\begin{abstract}
We report results of an extensive observational campaign 
of the 6$\,$cm formaldehyde maser in the young massive 
stellar object IRAS$\,$18566+0408 (G37.55+0.20) conducted
from 2002 to 2009.
Using Arecibo, VLA, and GBT, we discovered quasi-periodic 
formaldehyde flares (P $\sim$ 237 days).
Based on Arecibo observations, we also discovered correlated 
variability between formaldehyde (H$_2$CO) and methanol (CH$_3$OH) masers.
The H$_2$CO and CH$_3$OH masers are not spatially coincident, 
as demonstrated by different line velocities and high angular
resolution MERLIN observations. The flares could be caused by variations
in the infrared radiation field, possibly modulated by periodic
accretion onto a young binary system.
\end{abstract}


\section{Introduction}

\renewcommand{\thefootnote}{\arabic{footnote}}\setcounter{footnote}{0}

IRAS$\,$18566+0408 (G37.55+0.20) is a massive star forming region located
at a kinematic distance of 6.7$\,$kpc (Araya et al. 2004). 
The region is visible in Spitzer GLIMPSE infrared data 
(Benjamin et al. 2003), and shows an outflow
that has been detected in CO, SiO and in the 4.5$\,\mu$m 
IRAC band (Araya et al. 2007c; Beuther et al. 2002;
Zhang et al. 2007). At smaller scales, Very Large Array 
(VLA, NRAO)\footnote{The National Radio Astronomy Observatory 
is a facility of the National Science Foundation operated under 
cooperative agreement by Associated Universities, Inc.} 
observations reveal a radio continuum source, possibly an
ionized jet (Araya et al. 2007c). The bolometric luminosity of the object is 
$\sim 6 \times 10^4\,$L$_\odot$ (equivalent to an 
O8 ZAMS star; Zhang et al. 2007; Sridharan et al. 2002).

IRAS$\,$18566+0408 harbors one of the few known 6$\,$cm formaldehyde 
(H$_2$CO) masers in the Galaxy (Araya et al. 2008).
The H$_2$CO maser in IRAS$\,$18566+0408 showed a strong 
flare in 2002 (Araya et al. 2007b).
Here we report results of a monitoring program of
the H$_2$CO maser intended to investigate the
nature of the flare.

\section{Observations}

\subsection{Arecibo Observations}

Using the 305$\,$m Arecibo Telescope\footnote{The Arecibo 
Observatory is part of the National Astronomy and Ionosphere Center, 
which is operated by Cornell University under a cooperative 
agreement with the National Science Foundation.}, 
we monitored the 6$\,$cm H$_2$CO maser 
($J_{K_aK_c} = 1_{10} - 1_{11}$, 
$\nu_o = 4829.6594\,$MHz) in IRAS$\,$18566+0408.
The pointing position was R.A. = 18\h59\m09.98\sec, 
decl. = 04\arcdeg12\arcmin15.6\arcsec (J2000; Araya et al. 2005).
The monitoring program had three intervals:
May 2006 to April 2007, January 2008 to May 2008,
and October 2008 to November 2009. A total of 48 runs were conducted
(Figure~1; see Table~1, online version). 

We observed in position switching mode;
5 minutes on-source integration per scan with typically one or
two scans per run. Observations prior to June 2008 were conducted
with the Interim autocorrelator, dual linear polarization, 
1.56 MHz bandwidth and 2048 channels (0.047\kms~per channel). 
The data reduction was done using the CLASS program.\footnote{CLASS is 
part of the GILDAS software package developed by IRAM.} 
The WAPP spectrometer was used since October 2008, 
with 3.125$\,$MHz bandwidth and 2048 channels (0.095\kms~channel width).
The calibration of the WAPP data was done in IDL using 
Arecibo-IDL routines. The data were smoothed to a 
channel width of 0.19\kms.
We observed B1857+129 in most of the 
runs to check the pointing and measure the telescope gain. 
The pointing error was less than 12\arcsec~in general, and the 
gain varied between 5 and 9 K/Jy in the 48 runs. 
The system temperature was 
typically 26$\,$K. The Arecibo half power beam width (HPBW) is 
$\simeq 1$\arcmin~at 4860$\,$MHz.
We detected no significant variability of 
the linewidth or peak velocity of the maser.

The 6.7$\,$GHz CH$_3$OH  line ($J_k = 5_1 - 6_0 A^+$, 
$\nu_o = 6668.5192\,$MHz) was also observed with 
integration times on-source between 1 and 2 minutes. The same
Interim and WAPP configurations described above were
used for the CH$_3$OH observations. 
The spectra were smoothed to a channel width of 
0.07\kms~(see Figure~2).
The Arecibo HPBW is $\simeq 0.72$\arcmin~at 
6600$\,$MHz. A detailed description
of all CH$_3$OH observations together with results from an 
ongoing monitoring program will be reported in the
future.

\vspace{0.6cm}
\subsection{VLA Observations}

We used the VLA
to observe the 6$\,$cm H$_2$CO maser in IRAS$\,$18566 +0408
at seven epochs after 2006 (see Table~1, online version).
The pointing positions were 
R.A. = 18\h59\m10.10\sec, decl. = 04\arcdeg12\arcmin12.0\arcsec (J2000)
for the A and B array observations, and 
R.A. = 18\h59\m10.00\sec, decl. = 04\arcdeg12\arcmin25.0\arcsec (J2000)
for the C array observations. 
The seven observations, 
together with two previous VLA observations from the 
literature (Araya et al. 2007b), are shown in Figure~1. We used 
a 2IF mode with bandwidth of 1.56$\,$MHz 
(97$\,$km$\,$s$^{-1}$) and 255 channels 
(6.104$\,$kHz, 0.38$\,$km$\,$s$^{-1}$ channel width).
We observed 3C286 and 3C48 as flux density calibrators, with
assumed flux densities of 7.52$\,$Jy and 5.47$\,$Jy, respectively. 
We observed J1824+107 for amplitude and phase calibration.
We measured flux densities between 0.74 and 
0.79$\,$Jy for J1824+107 in the
seven runs, i.e., a dispersion of $\simeq 6\%$ which includes 
flux density calibration uncertainties and intrinsic variability
of the quasar. All data reduction 
was done with the NRAO software package AIPS following
standard spectral-line procedures.
The H$_2$CO maser positions measured with the VLA (A configuration) 
in 2003 and 2007 agree within $50\,$mas.

\subsection{GBT Observations}

The Green Bank Telescope (GBT) was used on five 
dates in 2008 to observe the H$_2$CO
maser (see Table~1, online version). 
The GBT H$_2$CO flux density measurements from 2008, together
with prior GBT observations from the literature (Araya et al. 2007b) are 
shown in Figure~1. 

The quasar J1851+005 was observed 
for pointing and focus adjustments.
The pointing corrections were less than 8\arcsec.
The observations of April 07 and 15, 2008,
were done in standard position switching
mode with integration times on-source between 
5 to 10$\,$min. 
The observations in May and October
2008 were conducted in frequency switching mode,
with integration time on-source between 10 and 30$\,$min.
We used the GBT spectrometer, 
with a bandwidth of 12.5$\,$MHz (775$\,$km$\,$s$^{-1}$),
8192 channels (0.094$\,$km$\,$s$^{-1}$ channel width)
and dual circular polarization.
The system temperature was $\simeq 20\,$K in all runs.
After checking for RFI and 
consistency between the two circular polarizations, 
the data were averaged and smoothed to a
channel width of 0.38$\,$km$\,$s$^{-1}$.  
All data reduction was done in IDL using the GBTIDL 
procedures.\footnote{http://gbtidl.nrao.edu/}

\subsection{MERLIN Observations}

High angular resolution observations of the 6.7$\,$GHz CH$_3$OH 
masers in IRAS$\,$18566+0408 were conducted
with Multi-Element Radio Linked Interferometer 
Network (MERLIN)\footnote{MERLIN is a National Facility operated 
by the University of Manchester at Jodrell Bank Observatory 
on behalf of STFC.} on April 03, 04, and 20, 2008, to determine
the location of the CH$_3$OH masers with respect to the
H$_2$CO maser. The pointing position 
was R.A. = 18\h59\m09.98\sec, decl. = 04\arcdeg12\arcmin15.6\arcsec (J2000).
The CH$_3$OH 
masers were observed with a bandwidth of 0.5$\,$MHz and 
255 channels (1.96$\,$kHz, 0.088\kms~channel width).
Each observing run lasted approximately 10 hours.
The quasar 1904+013 was observed as complex gain calibrator
with a broad bandwidth (13$\,$MHz) during approximately 2 minutes
for every $\simeq 7$ minutes observation of the CH$_3$OH masers.
We observed 3C84 in the narrow (0.5$\,$MHz) and wide 
(13$\,$MHz) setups to determine the solutions needed to
transfer the phase calibration from the wide 
to the narrow band. 3C84 was also used as bandpass 
and flux density calibrator (a flux density of 15.7$\,$Jy was assumed).

Poor phase tracking affected the observations of April 03 and
04, 2008, thus the astrometry of the MERLIN observations
discussed in this paper comes from the phase referenced 
observations of April 20, 2008.
To increase signal to noise, we combined all MERLIN data
and used the position of the brightest CH$_3$OH maser 
from the 2008 April 20 observations as initial 
model for self-calibration. The synthesized beam
of the self-calibrated data is 73 $\times$ 42$\,$mas, 
22\arcdeg~position angle; the final rms is $\simeq 12$\mJyb.
The self-calibrated data were used to create Figure~3. 
Including uncertainties in the phase reference source position,
the telescope positions, and phase transfer errors, we estimate
that the total astrometric uncertainty of our MERLIN data is 35$\,$mas.

\vspace{-0.9cm}
\section{Results}

We detected five new flare events and discovered
periodicity (see Figure~1); IRAS$\,$18566 +0408 is the 
only (quasi)periodic H$_2$CO maser system known. 
The flares have a periodicity of $\sim$237 days, a duration (at 
half-maximum) of $\sim$30 days, and an order of magnitude increase in
flux density. The peak flare intensity and maser intensity 
during the quiescent state show a monotonic decay (Figure~1). 
Possible causes of the decay include changes in
maser pumping and gas motions affecting beaming.

The flares are not strictly periodic. 
The period derived from the December 2006
and November 2008 flares is 238$\pm$4 days, and the period derived
from the December 2006 and March 2008 flares is 234$\pm$1 days.
The flare of August 2009 was well-sampled during the rising part of the
flare, but we do not have a good determination of its peak. 
Assuming that the measurement of August 15, 2009 corresponds to the
maximum of the flare, then the period would be 244$\pm$4 days.
If the 2002 flare were part of a periodic cycle of outbursts, 
then the period would be 248$\pm$5 days (the uncertainties
are 3$\sigma$ errors from autocorrelation analysis).
Most of the recent flares have a shorter period than the one 
estimated from the data prior to 2007.

The combined effect of (quasi)periodic flares, monotonic 
flux density decay, and tendency toward decreasing 
period as a function of time
are exemplified by the dotted-line in Figure~1. The dotted-line
was generated as follows: 1) the light-curve from May
2006 to January 2007 was used as a template; 
2) the template was
extrapolated assuming an exponential decrease in the interval
between consecutive flare peaks ($\tau = 237 \cdot 2^{-j/62}\,$days,
where $j = 1$ for the interval between the December 2006 and 
August 2007 flares, $j = 2$ for the interval between the August 2007 and
March 2008 flares, etc.), and an exponential decay in the maser flux density
($S_\nu \varpropto 2^{-(t-t_{\mathrm{ref}})/878}\,$mJy,
where $t_{\mathrm{ref}}$ is the Julian date of the December 2006 flare peak). 
This simple extrapolation (representative fit) 
based on the 2006 -- 2007 observations 
roughly reproduces the peak date and variability profile 
of the flares, as well as the flux density measurements 
between 2003 and 2005. The flare of August 2009 shows
that the period is neither constant nor strictly monotonically
decreasing. Nevertheless, the occurrence of the flares has been
regular enough to successfully predict the approximate
date of the 2008 and 2009 flares to schedule observations. Indeed, the
MERLIN observations were scheduled to coincide with the predicted 
flare of April 2008 (Figure~2).

We also monitored the 6.7$\,$GHz CH$_3$OH transition 
with Arecibo. The 6.7$\,$GHz CH$_3$OH  spectrum shows a group of
nine maser components spread between 78 and 
88$\,$km$\,$s$^{-1}$. Figure~2 shows the H$_2$CO
and CH$_3$OH spectra obtained on December 14, 2006, 
and the light curve of the H$_2$CO
maser and three CH$_3$OH maser components.
We discovered that the CH$_3$OH maser component 
at 87.8$\,$km$\,$s$^{-1}$ (hereafter component 9) 
shows flares that are similar to 
the H$_2$CO maser flares. Specifically,
the peaks of all CH$_3$OH (component 9) flares were 
simultaneous (within 10 days) to the peaks of the H$_2$CO 
flares.\footnote{The 2008 March flare is a possible exception. 
Two CH$_3$OH measurements separated by 14$\,$days gave
the same flux density within 3$\sigma$, which 
implies an uncertainly on the peak flare of $\sim$14 days.}
IRAS$\,$18566+0408 is thus
the first system where (quasi)periodic correlated maser
flares of CH$_3$OH {\it and} H$_2$CO molecules have
been detected, and one of only a few CH$_3$OH
periodic flare systems known (e.g., Goedhart et al. 2004, 2009).

The other CH$_3$OH maser components also show variability but not 
as well-correlated to the H$_2$CO maser (see for example
Figure~2, bottom panel). We discuss briefly the implications of the variability 
of the other CH$_3$OH masers in $\S$4, but an in-depth discussion
of the variability of all CH$_3$OH masers is beyond the scope of this
work.

\vspace{-0.6cm}
\section{Discussion}

The velocities of the CH$_3$OH maser component 9
and the H$_2$CO maser (Figure~2, insets) differ by
8.3$\,$km$\,$s$^{-1}$, and the masers 
are separated by $\sim$2000$\,$AU (12 light-days, 0.32\arcsec)
in projection (Figure~3). 
Given that the two masers showed simultaneous 
flares, the origin of the flares is not located in
either one of the maser regions, 
otherwise one component would have always flared
before the other. The different variability 
behavior of the different CH$_3$OH components shows
that the flares are not caused by an homogeneous
and large-scale change in the background radio continuum, 
e.g., in the flare of 2006, the CH$_3$OH component 3
did not show a flare even though it is 
located in between H$_2$CO and CH$_3$OH masers that
showed a flare (compare the position of the masers shown 
in Figure~3 with the light curves of Figure~2). We 
therefore find unlikely that the flares are caused
by a change in the background radio continuum.

It is also unlikely that the flares are triggered by a
propagating sound/density wave or shock front. 
For example, assuming that the CH$_3$OH component 9 and H$_2$CO masers
are located at the same distance from the source that 
triggers the variability (explaining simultaneous flares), then the 
minimum distance between the driving source and the masers
is $\sim 1000\,$AU (i.e., masers and driving source located in the plane
of the sky; see Figure~3).
This minimum distance implies that the travel time of a 
10$\,$km$\,$s$^{-1}$ wave from its origin
to the maser regions is $\gtrsim 500\,$yr,
which is far greater than the time scale of the flares
(i.e., $\sim$30 days flare duration, $\sim$237 days periodicity).
Faster (J-)shocks (100$\,$km$\,$s$^{-1}$ and greater) 
are unlikely because the molecules would disassociate 
with the passing of a single shock front (e.g., Hollenbach \& McKee 1989,
Garay et al. 2002), instead 
of showing periodic variability. The absence of significant 
changes in line peak velocity and linewidth, as well as
the smooth decay of the maser emission during the quiescent
phase argue against the J-shock hypothesis.

In contrast, a radiative origin of the flares appears
more likely. If the two maser regions
are in the plane of the sky with the source of 
pumping radiation between them (assumed hereafter to be 
the massive protostar or its surroundings),
then the radiation front would take only about
6 days to reach the maser regions. This
travel time provides a better match to the 
time scales of the flares.
In addition, changes in maser gain due to variability 
of the radiation field are consistent
with the excitation mechanism of Class II
CH$_3$OH masers (i.e., infrared pumping, Cragg et al. 2005). 

Given that the CH$_3$OH and H$_2$CO masers show correlated
variability, the excitation mechanism of the two 
maser species must be similar. The excitation mechanism
of H$_2$CO masers has been a controversial topic
(Araya et al. 2006, 2007c; Hoffman et al. 2003); 
our discovery of correlated H$_2$CO and CH$_3$OH
masers shows that the excitation mechanism
of H$_2$CO masers is probably infrared pumping.

The periodic behavior of the 
flares could be caused by a variety of astrophysical 
processes; chief among these are stellar pulsations and binary induced
variability. 
Some types of pulsating variable stars have periods of hundreds
of days (e.g., Eyer et al. 2008). Nevertheless, the light 
curves of such pulsating stars are significantly 
different from the flare-like periodic variability of
the H$_2$CO and CH$_3$OH masers. Pulsations of massive 
protostars (and/or inner accretion disks) cannot be ruled 
out until detailed numerical modeling is conducted.

An alternative possibility appears more likely:
variability modulated by a binary system.
Binarity and multiplicity are ubiquitous characteristics of 
massive stars on both theoretical and observational grounds
(e.g., Krumholz et al. 2009; Kraus et al. 2007).
The orbital periods of visual and spectroscopic
massive binaries range between a few days 
to many years (e.g., Zinnecker \& Yorke 2007),
encompassing the time scale of the 
periodic maser flares in IRAS$\,$18566+0408. In addition,
the observed misalignment between an ionized jet,
a {\it Spitzer} IRAC 4.5$\,\mu$m excess, and
a SiO outflow could 
be caused by precession due to a binary system
(Araya et al. 2007c, Zhang et al. 2007).

Recently, van der Walt et al. (2009) proposed that the CH$_3$OH
flares in G9.62+0.20E (which have a periodicity very similar to 
the H$_2$CO and CH$_3$OH flares in IRAS$\,$18566+0408) 
are modulated by variability from a colliding wind
binary, responsible for a change in the background radio
continuum and/or pumping radiation.
Even though such a colliding wind model reproduces
the variability of the CH$_3$OH maser flares in G9.62+0.20E,
we consider the model unlikely in the case of 
IRAS$\,$18566+0408. The H$_2$CO flares are not strictly 
periodic (Figure~1), thus, an additional, stochiastic mechanism 
is implicated. Minor local turbulence
could superimpose irregularities on the response to a periodic cause but
does not explain the combination of decay and irregularity.

Here we propose an alternative scenario for the 
maser flares in IRAS$\,$18566+0408:
periodic accretion of circumbinary disk material. 
This process has been predicted  
and has observational evidence in the case of 
some young, low mass binaries (Artymowicz \& Lubow 1996;
Jensen et al. 2007; G\"unther \& Kley 2002; Mundt et al. 2010).
In this scenario, material 
from the circumbinary disk is accreted onto the protostars or
accretion disks, heating the dust and increasing the infrared 
radiation field, resulting in higher microwave amplification due 
to greater maser gain. 
For example, the smoothed particle hydrodynamics simulations 
by Artymowicz \& Lubow (1996) show that a binary system with mass ratio 0.79 
and eccentricity $e = 0.5$, will experience bursts of accretion onto the 
binary components with a dimensionless time-dependence that is
quite similar to the H$_2$CO light-curve.
In a binary with mass ratio $\sim$0.8 (e.g., a 20 and 
16\msol binary) and $e = 0.5$, a periodicity of 
$\sim$240$\,$days is expected if the semimajor axis 
of the most massive component is $\sim$1.1 AU.
The short flare events traced by H$_2$CO and CH$_3$OH
masers in IRAS$\,$18566+0408 (this work) and CH$_3$OH masers in 
G9.62+0.20E (Goedhart et al. 2004) could  be caused 
by an orbital configuration similar to the one discussed 
above. Less flare-like (more undulated and/or aperiodic) 
variability seen in other CH$_3$OH maser sources (Goedhart et al. 2004) 
would be expected from
binary systems with lower eccentricities (see Artymowicz \& Lubow 1996).
This is, the model of Artymowicz \& Lubow (1996) could
explain not only the time scale for the periodicity, but also the
duration of the flaring events, depending essentially on the
eccentricity of the binary system.
Thus, periodic maser flares could trace 
the properties of young massive binaries that are 
actively accreting. Moreover, accretion of circumbinary material 
may cause mass equalization in massive 
binaries. For example, in the model of Artymowicz \& Lubow (1996),
the lower mass component of the binary is the one with a higher 
accretion rate (see however G\"unther \& Kley 2002).

\section{Summary}

We detected the first (quasi)periodic H$_2$CO maser flare
system. The maser is coincident with a massive protostellar candidate
(IRAS$\,$18566+0408). The periodicity of the flares is 
approximately 237$\,$days. We also detected 6.7$\,$GHz CH$_3$OH flares
that are correlated to the H$_2$CO outbursts. Regardless of
whether the H$_2$CO maser flares reported in this work 
unveil tight massive binaries still undergoing accretion or 
some other process,
our discovery shows that short time-scale changes (weeks to months)
in physical conditions surrounding 
massive protostars are not random, but that 
underlying (semi)harmonic mechanisms are at work 
during the process of massive star formation.

\acknowledgments

This work was partially supported by a Jansky Fellowship of the NRAO.
P.H. acknowledges partial support from NSF grant AST-0908901.
We thank the scheduler officers of NAIC, NRAO, and MERLIN 
for their flexibility on scheduling the observations reported here, in 
particular H. Hern\'andez at Arecibo. We also thank an anonymous 
referee for comments and suggestions that significantly improved the
manuscript.

\clearpage



\clearpage

\begin{figure}
\includegraphics{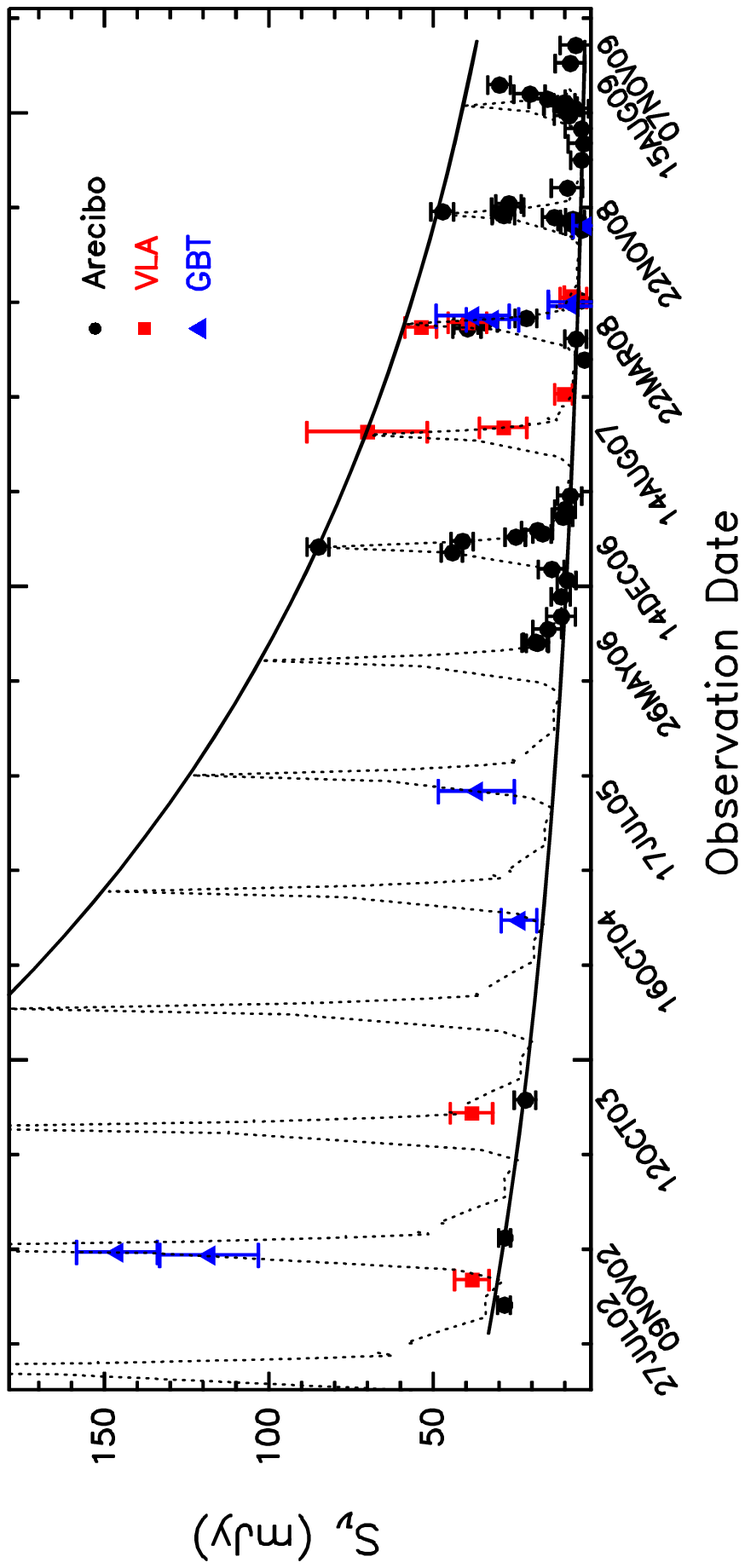} 
\vspace*{12cm}\caption{Light curve of the H$_2$CO maser flux density
in IRAS$\,$18566+0408. Data obtained with Arecibo, VLA 
and GBT are shown
with different symbols. Measurements before 2006 are from 
Araya et al. (2007b). The dotted line is an extrapolation of the
2006-2007 Arecibo observations assuming an exponential decay 
of the flux density and a slow decrease
of the flare period (see $\S$3). The solid lines are the functions
$S_{\nu,i} = A_i\,2^{-(t-t_{\mathrm{ref}})/878}\,$mJy, where 
$t_{\mathrm{ref}} = 2454084\,$JD (the 
date of the December 2006 flare peak), and $A_i$ is 85 and 9.0 for the upper
and lower curves, respectively.}
\label{f2}
\end{figure}

\clearpage

\begin{figure}
\includegraphics{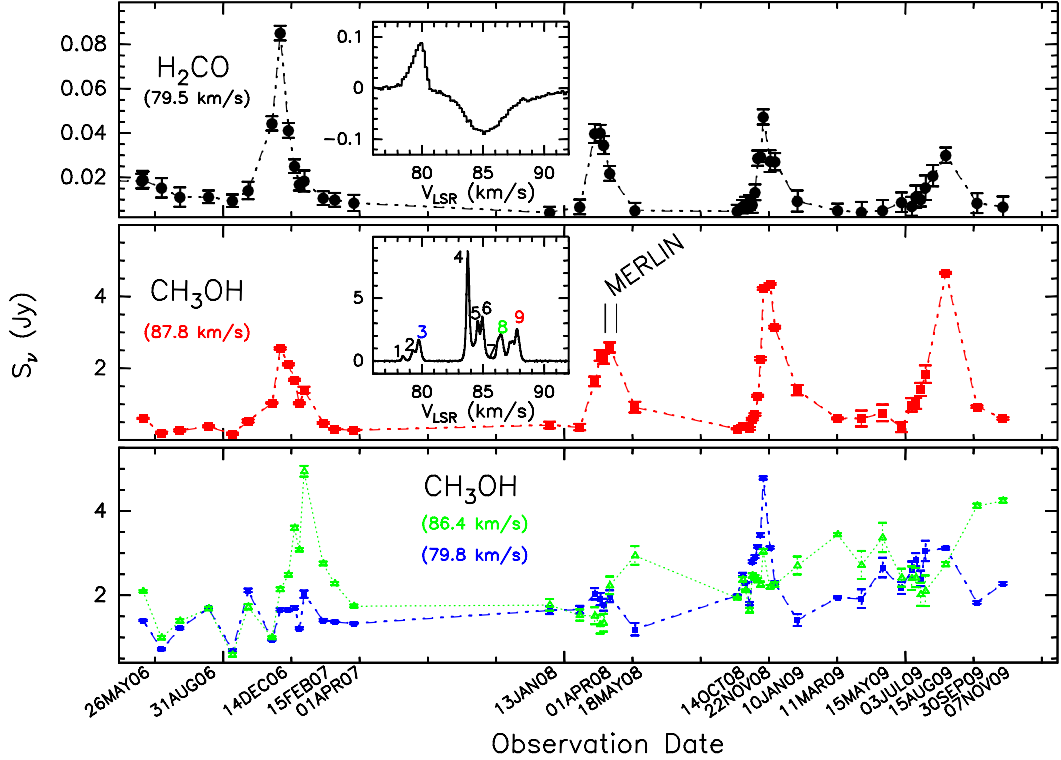} 
\vspace*{11cm}\caption{Spectra of H$_2$CO and CH$_3$OH masers 
obtained with the Arecibo Telescope on December 14, 2006 (insets). 
We show the Arecibo Telescope light curves from May 2006 to 
November 2009 of the H$_2$CO (top panel) and of three 
CH$_3$OH maser components (middle and bottom panels).
The H$_2$CO absorption
at 85$\,$km$\,$s$^{-1}$ originates from the extended molecular 
cloud where the massive (proto)star is located. The dates of the 
MERLIN observations (Figure~3) are marked with vertical lines.}
\label{f3}
\end{figure}

\clearpage

\begin{figure}
\includegraphics{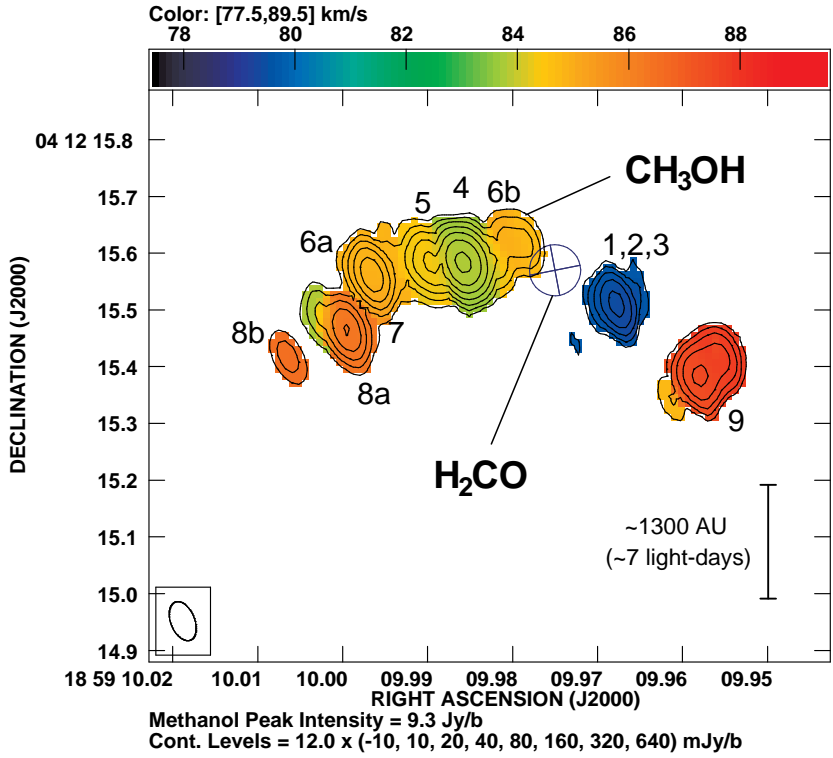} 
\vspace*{12cm}\caption{Velocity field of 6.7$\,$GHz CH$_3$OH 
masers in IRAS$\,$18566+0408 obtained with MERLIN 
($\theta_{syn} = 72 \times 42\,$mas, Position angle = 22$^o$; April 2008). 
The labelled components correspond to the spectral
features marked in Figure~2 (inset, lower panel), 
and the variability curve of features 3, 8, and 9 are shown in Figure~2 
(lower and middle panels).
The ``$\oplus$'' symbol shows the location of the 6$\,$cm H$_2$CO  
maser ($\theta_{syn} = 0.5$\arcsec $\times$ 0.4\arcsec, 
P.A. = 11\arcdeg, VLA observations, Araya et al. 2005).
The size of the H$_2$CO symbol represents 3$\sigma$ positional 
uncertainty from the fit.} \label{f4}
\end{figure}

\newpage

\begin{deluxetable}{rccl}
\tablewidth{0pt}
\tablecaption{H$_2$CO Monitoring Observations}
\tablehead{
\colhead{Date}              &
\colhead{S$_{\nu}$ (mJy)}    &
\colhead{RMS (mJy)}         &
\colhead{Telescope}}
\startdata
27 Jul. 2002  &   28.5      &      1.3    &  Arecibo          \\ 
18 Sep. 2002  &   38.3      &      3.5    &  VLA-CnB          \\ 
09 Nov. 2002  &   118       &      10     &  GBT              \\ 
15 Nov. 2002  &   146.2     &      8.2    &  GBT              \\ 
16 Dec. 2002  &   28.3      &      1.2    &  Arecibo          \\ 
05 Sep. 2003  &   38.4      &      4.3    &  VLA-A            \\ 
12 Oct. 2003  &   22.1      &      2.2    &  Arecibo          \\ 
16 Oct. 2004  &   23.9      &      3.6    &  GBT              \\ 
17 Jul. 2005  &   36.9      &      7.7    &  GBT              \\ 
25 May  2006  &   18.4      &      2.2    &  Arecibo          \\ 
26 May  2006  &   19.0      &      2.7    &  Arecibo          \\ 
27 May  2006  &   19.0      &      2.4    &  Arecibo          \\ 
23 Jun. 2006  &   15.3      &      2.9    &  Arecibo          \\ 
20 Jul. 2006  &   11.2      &      2.9    &  Arecibo          \\ 
31 Aug. 2006  &   11.2      &      1.9    &  Arecibo          \\ 
05 Oct. 2006  &    9.5      &      1.9    &  Arecibo          \\ 
28 Oct. 2006  &   14.1      &      2.6    &  Arecibo          \\ 
02 Dec. 2006  &   44.3      &      2.2    &  Arecibo          \\ 
14 Dec. 2006  &   85.1      &      2.2    &  Arecibo          \\ 
26 Dec. 2006  &   41.2      &      2.2    &  Arecibo          \\ 
04 Jan. 2007  &   25.1      &      2.1    &  Arecibo          \\ 
11 Jan. 2007  &   16.9      &      1.9    &  Arecibo          \\ 
18 Jan. 2007  &   18.4      &      3.1    &  Arecibo          \\ 
15 Feb. 2007  &   10.7      &      2.1    &  Arecibo          \\ 
04 Mar. 2007  &    9.9      &      2.1    &  Arecibo          \\ 
01 Apr. 2007  &    8.5      &      2.5    &  Arecibo          \\ 
14 Aug. 2007  &   70        &      12     &  VLA-A            \\ 
24 Aug. 2007  &   28.7      &      4.8    &  VLA-A            \\ 
01 Nov. 2007  &   10.4      &      1.8    &  VLA-B            \\ 
13 Jan. 2008  &    4.2      &      1.8    &  Arecibo          \\ 
26 Feb. 2008  &    6.7      &      2.2    &  Arecibo          \\ 
19 Mar. 2008  &  39.8       &      2.9    &  Arecibo            \\ 
22 Mar. 2008  &  53.8       &      3.2    &  VLA-C              \\ 
27 Mar. 2008  &  40.0       &      2.5    &  Arecibo            \\ 
01 Apr. 2008  &  39.6       &      3.9    &  VLA-C              \\ 
01 Apr. 2008  &  34.5       &      2.7    &  Arecibo            \\ 
07 Apr. 2008  &  32.0       &      5.3    &  GBT                \\ 
11 Apr. 2008  &  21.8       &      2.2    &  Arecibo            \\ 
15 Apr. 2008  &  38.0       &      7.4    &  GBT                \\ 
05 May  2008  &  $<15$      &      5.0    &  GBT                \\
14 May  2008  &  $<15$      &      5.0    &  GBT                \\
18 May  2008  &   5.1       &      2.3    &  Arecibo            \\ 
23 May  2008  &   7.4       &      2.7    &  VLA-C              \\ 
26 May  2008  &   7.9       &      1.5    &  VLA-C              \\ 
14 Oct. 2008  &   4.8       &      1.9    &  Arecibo            \\ 
22 Oct. 2008  &  $<7.5$     &      2.5    &  GBT                \\
23 Oct. 2008  &   6.8       &      2.0    &  Arecibo            \\ 
27 Oct. 2008  &   7.9       &      2.2    &  Arecibo            \\ 
02 Nov. 2008  &   8.7       &      2.2    &  Arecibo            \\ 
06 Nov. 2008  &   7.7       &      2.4    &  Arecibo            \\ 
10 Nov. 2008  &  13.3       &      2.4    &  Arecibo            \\ 
14 Nov. 2008  &  28.7       &      2.3    &  Arecibo            \\ 
18 Nov. 2008  &  29.4       &      1.7    &  Arecibo            \\ 
22 Nov. 2008  &  47.3       &      2.3    &  Arecibo            \\ 
02 Dec. 2008  &  27.4       &      3.3    &  Arecibo            \\ 
09 Dec. 2008  &  27.1       &      2.6    &  Arecibo            \\ 
10 Jan. 2009  &   9.4       &      3.2    &  Arecibo            \\ 
11 Mar. 2009  &   5.1       &      2.1    &  Arecibo            \\ 
14 Apr. 2009  &  $<9.3$     &      3.1    &  Arecibo            \\
15 May  2009  &  $<9.6$     &      3.2    &  Arecibo            \\
12 Jun. 2009  &   8.8       &      2.9    &  Arecibo            \\
27 Jun. 2009  &   7.0       &      2.9    &  Arecibo            \\
03 Jul. 2009  &  11.3       &      3.4    &  Arecibo            \\
09 Jul. 2009  &  10.1       &      2.1    &  Arecibo            \\
17 Jul. 2009  &  15.4       &      3.7    &  Arecibo            \\
27 Jul. 2009  &  20.8       &      3.2    &  Arecibo            \\
15 Aug. 2009  &  30.0       &      2.3    &  Arecibo            \\
30 Sep. 2009  &   8.4       &      3.0    &  Arecibo            \\
07 Nov. 2009  &   6.8       &      3.1    &  Arecibo            \\
\enddata
\tablecomments{
\small Data prior 2006 is from Araya et al. 
(2004, 2005, 2007a, 2008). The synthesized beams of our
VLA observations in A, B, and C configurations 
were approximately 0.5\arcsec, 1.7\arcsec, and
5\arcsec, respectively.}
\end{deluxetable}

\end{document}